\documentclass[bibtex,nonblind]{apsr_submission}

\title{Deep Active Learning for Data Mining from Conflict Text Corpora}

\author{Mihai Croicu}
       {Department of Peace and Conflict Research, Uppsala University and Peace Research Institute Oslo (PRIO)}

\thanks{Draft date: 2024-01-31. The author would like to thank Håvard Hegre, Simon von der Maase, Joakim Kreutz, Thorsten Rogall and the participants to the 2024 Using LLMs and Text-as-Data in Political Science Research Workshop in Barcelona. The research
was funded by the European Research Council, grant number 101055176 - ANTICIPATE and by Riksbanken Jubileumsfond grant number M21-0002 - Societies at Risk. Computation was carried out on the Alvis cluster of the National Academic Infrastructure for Supercomputing in Sweden, grant number NAISS 2023/5-551.}

\runningtitle{Deep Active Learning for Data Mining from Conflict Text Corpora}
\runningauthor{Croicu} 

\usepackage{amsmath}
\usepackage{graphicx}
\usepackage{rotating}

\usepackage{pdflscape}
\usepackage{subfig}
\usepackage{multicol}
\usepackage{multirow}

\begin{document}
\begin{frontmatter}
\begin{abstract}
High-resolution event data on armed conflict and related processes have revolutionized the study of political contention with datasets like UCDP GED, ACLED etc. However, most of these datasets limit themselves to collecting spatio-temporal (high-resolution) and intensity data. Information on dynamics, such as targets, tactics, purposes etc. are rarely collected owing to the extreme workload of collecting data. However, most datasets rely on a rich corpus of textual data allowing further mining of further information connected to each event. This paper proposes one such approach that is inexpensive and high performance, leveraging active learning - an iterative process of improving a machine learning model based on sequential (guided) human input. Active learning is employed to then step-wise train (fine-tuning) of a large, encoder-only language model adapted for extracting sub-classes of events relating to conflict dynamics. The approach shows performance similar to human (gold-standard) coding while reducing the amount of required human annotation by as much as 99\%.
\end{abstract}
\end{frontmatter}

\section{Introduction and Motivation}

The development of incident / battle event level datasets on armed conflict such as the Uppsala Conflict Data Program's Georeferenced Event Dataset \citep{sundberg2013introducing}, ACLED \citep{raleigh2010introducing} or SCAD \citep{salehyan2012social} have revolutionized research in the micro-dynamics of armed conflict in the past decade, allowing the exploration of research agendas as ranging from as diverse topics as the climate-conflict nexus \citep{von2014sustained} to the local effects of peacekeeping deployments \citep{ljungkvist2021revisiting} and allowed, for the first time, for forecasting of conflict dynamics at a local, disaggregated level \citep{hegre2022lessons}.

However, one significant limitation of all such datasets is the amount of features they contain -- usually limited to geography, time, participating actors (usually at an aggregate level), some measures of event intensity and a coarse classification (such as whether a conflict involved the state, or whether a protest was violent or not). These limitations are caused by the extreme efforts required to collect such data using manual labor -- usually in the range of person-decades of work for every year covered \citep{hegre2020introducing}.

This, however, limits the types research questions that can be studied with such micro-level data, especially in large-scale, generalizable, comparative contexts, since data is simply lacking. Exploration of micro-dynamics of conflicts (e.g. territorial control, base formation, movement), targeting patterns (e.g. religious violence, attacks on infrastructure, urban and rural patterns), tactics (armaments used, formations etc.), while underlying the theoretical underpinning of conflict research \citep{kalyvas2006logic} is thus limited in scope and nature.

One approach employed to round this problem is the now-standard quantitative single-case research approach \citep{thaler2017mixed}, that leverage smaller, usually local data collection efforts, frequently based on surveys, with much richer feature sets, but typically limited to a single country, in order to draw inference on various patterns of conflict dynamics. 

These are too many to exhaustively list across all domains -- but even for one of the applications presented in this paper -- electoral violence, they are incredibly rich in terms of the details of data collection. These include explorations of local drivers and causes (incidental versus determinate) of electoral violence in C\^ote D'Ivoire \citep{van2023patterns}, the interplay between rebel group fragmentation, polarization and presence and the resulting patterns of electoral violence in Burundi \citep{colombo2019rebellion}, the role of kinship networks in the transmission chains of electoral violence in Nigeria \citep{fafchamps2013political}, the role of vote-buying in electoral violence \citep{bratton2008vote} to name but a few.

Attempts at generalization have been, however, thwarted by the lack of availability of data. This meant that going beyond a single case (or a small universe of related cases) forced the researcher to rely on extremely coarse (usually country-year) aggregate covariates, frequently only being able to proxy the quantities of interest with coarser measures, such as in \citet{brancati2013time} or \citet{fjelde2016electoral}. While this research is essential, the coarseness of the data, as well as the usage of datasets that are collected for a different purpose can lead to measurement errors and biases \citep{von2021systematic}.

Thus, obtaining good data on micro-dynamics of conflict is essential. This can be shown with ease, since, recently, some large-N cross-national datasets on conflict micro dynamics did begin to be collected and their impact can be seen almost immediately. As such, as soon as datasets such as the Deadly Electoral Conflict Dataset \citep{fjelde2022introducing} appeared -- studies employed them in conjunction with complex identification strategies to test the generalization potential of previous arguments were performed in e.g. \citet{fjelde2022protecting} or \citet{smidt2021keeping} with findings bridging previously unassailable gaps between multiple literature strands (e.g. the relationships between peacekeeper intervention and electoral conflict).

However, collecting such data is extremely expensive in terms of resources -- usually requiring a multi-year, large staff research project for even a single domain -- such as electoral violence \citep{fjelde2016electoral} or disaggregated conflict issues \citep{brosche2023they} or attacks on religious targets \citep{kruetz2018gods}. This entails that usually such datasets will collect a very limited amount of covariates for a very limited amount of subjects, and will frequently not be updated with a regularity that will make them available for e.g. forecasting efforts.

This paper presents a mostly automated approach to collect such micro-level data from the existing text-based corpora of large data collection projects such as the Uppsala Conflict Data Program (UCDP) in their Geo-referenced Event Dataset (GED) \citep{sundberg2013introducing}. 

This approach exploits four distinct methodological and data availability advancements in recent years:
\begin{enumerate}
    \item The vast annotated corpora of (battle-)event data collected by UCDP with global coverage and long time-series. These corpora, while only collecting limited amounts of features, are extremely large -- over 30 million words, and are entirely human curated, in e.g. filtering the relevant paragraphs of text.
    \item The development of datasets manually mining micro-data from the above corpora, allowing the development and testing of an automatic process to replicate this manual process with (much) more limited resources in terms of human annotation.
    \item Advances in modelling language computationally and the explosion of high-perforamce deep learning large language models (LLMs). These are advanced pretrained machine learning models that build their own representation of natural text from text, such as GPT and BERT and allow for much higher performance in creating a machine representation and understanding of human language that includes contextual and semantic relationships \citep{vaswani2017attention, brown2020language, devlin2018bert}.
    \item Developments in (mostly) theoretica computer science bringing together learning, where humans inform models interactively, while they are in their training phase, iteratively reducing uncertainty at each step of the training process \citep{dor2020active, burr2009}.
\end{enumerate}

and proposes a human-in-the-loop (active learning), deep-learning based machine learning system in order to mine the existing large conflict-related corpora for various classes and types of micro-dynamics of conflict with very limited human coding effort, reducing the need for human annotation by as much as 99\%.

The paper then validates this approach by attempting to recreate the main features of two gold-standard human coded datasets with these coding budgets - DECO \citep{fjelde2016electoral} and the Relgious Attacks Dataset \citep{kruetz2018gods}.

\section{Potential alternatives -- a short literature review}
One potential alternative to this solution would be to use a fully unsupervised dependency parsing approach, leveraging semantic relations between entities and actions that exist in the text in order to extract sets of known and pre-defined semantic patterns (e.g."subject actor - verb action - object actor"). This approach has a long history in the field, exemplified by automatic event coders such as TABARI, PETRARCH \citep{brandt_cameo} or their large language model (LLM) successors such as Multi-CoPED \citep{skorupa2022multi}. While these are extremely useful for many types of enquiries, they are not particularly suitable for information extraction from an already annotated corpus -- as these are extremely strict in their inclusion criteria (require actions to be labeled to belonging to a pair of known actors) whereas this task requires a more comprehensive inclusion criteria, i.e. a wider fishing net to select candidates, since a lot of information is present in the context rather than in the description of the action itself. For example, in the following text: \textit{"Two car bombs killed nine people in the Kurdish-held Iraqi oil city of Kirkuk on Thursday, police and medical sources said. The explosions near a Shi'ite mosque holding displaced people came after an offensive launched by Islamic State" (Reuters, 2014-08-07, Car bombs in Iraq's Kirkuk kills nine people)}, while this was an attack against a religious building, the building is only present as a location identifier, and not as a direct or indirect object; the context determines the nature of the attack. A strict dependency parser like PETRARCH, even with custom dictionaries, would have trouble identifying this as even a possible candidate. Indeed, the purpose of these methods is different to the task at bay -- they are wide, focusing on expanding knowledge outwards -- e.g. trying to explore the entire semantic space for networks across wide breadths of interlinked domains (such as mediation or conflict), not explore a single topic in depth. This continues to be the case with the most recent pushes in this field -- attempting to identify new actions (and compile new dictionaries of actors and actions -- nouns and verbs) to expand and complement existing ones to furthe refine the extracted networks in these broad fields \citep{haffner2023introducing, parolin2021come}.

A second approach would be to use a traditional machine learning approach for information extraction - where sufficient observations are labelled in advance by a human annotator and then a (single-shot) machine learning model is trained to recognize the same patterns as those identified by the human coder. This can be any kind of statistical (machine learning model) - either a classical shallow learner where features are extracted from text explicitly such as \citet{croicu2015improving} or \citet{rodriguez2022word}, a more modern representation learning approach where features are extracted automatically from text such as \citet{kristineNNN} or an even more advance approach such as \citet{laurer2022less} or \citet{zhang2019casm}, where text data is augmented with structural data in the form of known hypotheses about relationships present in the text.

The problem with such approaches is that they require a large corpus of annotated text to work, especially in highly imbalanced contexts such as e.g. electoral violence or religious violence or tactics or territorial control, where the subject of interest is very rare, perhaps appearing in 1--2\% of all the texts of the corpus. In such cases, simply sifting through the negatives to make a training dataset is an extremely time- and money- consuming effort, requiring the labelling of tens of thousands \citep{kristineNNN} or hundred of thousands of articles \citep{croicu2015improving} to reach acceptable performance, basically requiring the same level of work as in the case of manual labor.

\section{Validation datasets}

In order to evaluate the proposed method, the paper relies on two high quality, already annotated datasets that (manually) mine an existing conflict corpus for information.

Both datasets mine the Uppsala Conflict Data Program (UCDP) corpus that underlies the Georeferenced Event Dataset (GED)\citep{sundberg2013introducing} in order to extract information on conflict dynamics that are not present in the original dataset.

The first such dataset extracts information on attacks against religious institutions and clergy members \citep{kruetz2018gods}. The second such dataset is the Deadly Electoral Outcomes dataset (DECO) \citep{fjelde2022introducing}, capturing instances of violence against electoral infrastructure and processes.

Both datasets are, for the most part, manually mined with limited automatic pre-processing (such as simple keyword search for creating hierarchies of articles to code), and have extremely well developed coding methodologies and validation procedures to insure internal consistency \citep{kruetz2018gods, fjelde2022introducing}\footnote{In the case of the religious violence dataset, a formal inter-coder reliability test was performed, where three annotators independently coded 500 articles, with a Cohen's $k$ exceeding $.9$.}.

Furthermore, both datasets are, as expected, extremely unbalanced : DECO has 4,017 positives from a total of 187,577 coded events, whereas the religious attacks data captures 1,982 positives from a total of 193,114 events\footnote{The differing size of the corpora comes from the different base date considered by the datasets' author when initiating the coding process. For this approach, the exact corpus used by them was employed, for maximum validity of the method.}.

Each event represents one individual battle resulting in at least one fatality at a given date and for a given geo-referenced place. To each such event one or more source articles are attached -- a body of text, mostly news articles from global news-wires\footnote{These are Reuters, Agence France Presse, Associated Press, Xinhua, BBC sourced from Dow Jones Factiva using a fixed search string\citep{croicu2015ucdp}.} complemented with local NGO and other reports, from which the data in the event was extracted\footnote{A further explanation of UCDPs data collection methodology, sources etc. is outside the scope of the paper, but explained in detail in \citet{croicu2015ucdp}.}.

\section{An active learning approach}

Active learning is a subset of semi-supervised machine learning where the algorithm is able to interactively query a human coder (referred to as an oracle or annotator) during its learning process -- asking the human to iteratively validate and improve (uncertain) predictions made by the machine learning model -- and using these validated predictions, in order to refine its learning process \citep{zhan2021comparative, zhang2022survey, dor2020active, settles2007multiple, settles2011theories}. Essentially, it is an attempt to model traditional (human) seminar-style learning in a machine learning framework \citep{settles2007multiple, settles2011theories, burr2009} -- where questions are asked to the tutor during the learning session in order to address the most confusing or least certain parts.

Since active learning hones and refines the classifier during its training (optimization) period, this allows much better exploitation of the data \citep{zhan2021comparative, settles2007multiple} -- as, with a proper data selection strategy for querying the human annotator (referred to as the querying strategy \citep{dor2020active, settles2007multiple}, it allows the machine learning model to improve on specifically those cases it has the least understanding of, thus maximizing the improvement in the decision boundary. This allows for much less data than traditionally required to create efficient and effective machine learning models \citep{zhang2022survey}. As such, with only a limited amount of known (labelled data) one can exploit a much larger pool of unlabelled data efficiently \citep{zhan2021comparative}.

Active learning is seeing a resurgence in natural language processing research across a broad front, especially when coupled with deep learning approaches \citep{zhang2022survey, dor2020active}, specifically for two reasons --an increase in unlabelled or poorly labelled data that can be exploited in traditional approaches only with the associated large costs of annotation; and the improved efficiency of deep learning techniques to extract information from few samples. Further, active learning is making some inroads in social sciences -- especially when it comes to traditional text-based analyses like frame extraction or sentiment analysis in e.g. sociology \citep{bonikowski2022politics} where it has been employed successfully to e.g. extract narrative frames such as nostalgia.

The method consists of four different elements \citep{dor2020active, zhang2022survey}:
\begin{enumerate}
    \item \textbf{What's the end-goal?} - or \textbf{when to stop?} -- how does one best evaluate the model, what budgets for human annotations to employ and when is a model sufficiently improved to have been considered as converged on the final results?\citep{zhang2022survey, dor2020active}.
    \item \textbf{Where to start} - or how to generate an initial pool of labelled data for the machine learning model to begin learning \citep{dor2020active}. This is sometimes called pre-annotation \citep{skeppstedt2013annotating} and usually employs a simpler (or pre-existing) unsupervised model to rank the unlabelled data for the human to use.
    \item \textbf{How to learn?} - or what model best allows for modelling the relationship between the text and the information we want extracted \citep{zhang2022survey}
    \item \textbf{How to employ the human?} - also known as the querying strategy \citep{zhang2022survey, dor2020active}, what is the best way to select the small subset of uncertain cases that the human will validate. There are multiple approaches for this - either based on uncertainty (what observations the model is least certain about) or diversity (can the human be presented with a subset representing various aspects of the data space \citep{zhang2022survey, dor2020active}.
\end{enumerate}

\begin{figure}[htb!]
\includegraphics[width=9.9cm]{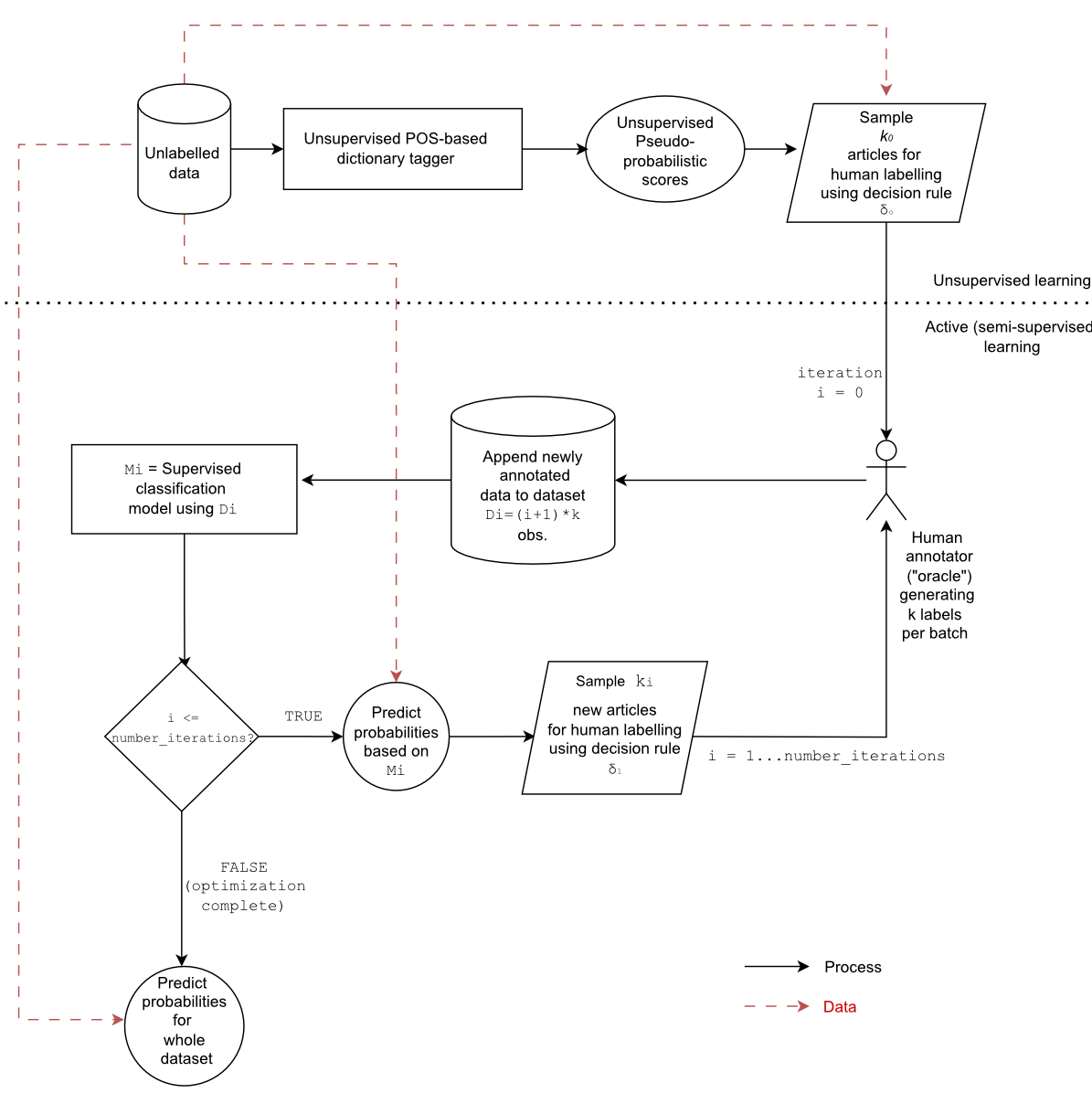}	
\caption{A schematic representation of the active learning approach employed in the paper. The top row of the figure describes the fully unsupervised step (the extraction of the "core" seed dataset for annotation). Below, the active learning iterative loop is presented. At this step, a model $M$ is iteratively improved through sequential training. This is done using batches of $k$ human annotated observations (news articles) that are selected from the core dataset for annotation by the previous iteration of $M$.} \label{fig:arch}
\end{figure}

The process itself is schematically described in figure \ref{fig:arch}, and can broadly be divided into two stages. In the first stage an unsupervised process tentatively filters the unlabeled dataset for increased balance by eliminating the most unlikely cases through the use of a keyword-based technique. In the second stage, an iterative batch active learning process is used, where a machine-learning classifier is iteratively improved by being step-wise trained on a small number of human-annotated examples that are selected based first on the results of the first stage and then on the results provided by the previous iteration of the classifier.

The main suggestions made by computer science literature, are, while diverse, connected by a red thread - the need to maximize the extraction of (true) minority class observations from the unlabeled set. This paper leverages a few of these suggestions. Firstly, it employs the selection of a smaller subset of observations as candidates for labeling with this subset being designed in such a way as to have a higher balance than the whole dataset \citep{ertekin2007learning}. Secondly, it employs a dictionary-based approach to build the previously-mentioned subset, using a keyword-search as an unsupervised seed for the active-learning process, in an approach similar to \citep{dor2020active}. Thirdly, during the active learning stage it iteratively constructs the set of observations by using a two-stage algorithm which in effect is a simplified version of \citet{lin2018}'s \texttt{MB-CB} algorithm: using the previous iteration of the classifier to extract a set of likely positives from the dataset for annotation, annotating them, and then, if balance is skewed, using the most unbalanced subset to extract unlabeled samples that are then treated as true negatives\footnote{This parallels the \texttt{LABEL-ACTIVE} and \texttt{ADD-RANDOM-NEG} approaches in \citet{lin2018}}.

\section{End goal: A note on model evaluation}

\subsection{Metrics}

The paper will employ two evaluation metrics for gauging the performance of the active learning approach - the area under the precision-recall curve (AUPR) or Average Precision (AP) \citep{sofaer2019area, davis2006relationship, boyd2013area} and the area under the receiver operating characteristic curve (AUROC) \citep{davis2006relationship, janssens2020reflection}. Both are composite metrics for probabilistic classifiers that describe their discriminating capacity. 

Both have become standard in predictive conflict research \citep{hegre2019views}, allowing for easier understanding of the performance of the approach in comparison with other models in the field.

AP shows the trade-off between precision (the proportion of elements labeled by the method as positive that were actually correct) and recall (the proportion of actual positives that were correctly identified) across the entire probability range \citep{boyd2013area, sofaer2019area}. In effect, AP sweeps over each possible threshold choice of predicted probability score between 0 and 1, uses that threshold to compute prediction and recall, plots these on a curve, and then computes the area under that curve. In a perfect scenario, of perfect separation and thus a perfect classifier, where irrespective of threshold, all true positives are extracted correctly, AP is 1, with a random model having an AP of 0 \citep{boyd2013area}. AP is usually computed as as a finite sum using the trapezoidal rule, taking each predicted probability of the classification as a threshold $t$ \citep{boyd2013area, sofaer2019area}:

$$\text{AP} = \sum_{t=1}^{N} (\text{Recall}_t - \text{Recall}_{t-1}) \times \text{Precision}_n$$

AUROC is a similar metric, but this time computing the trade-off between recall and false positive rates, i.e. the proportion of a true negative labelled as positives against the total number of negatives in the dataset \citep{hanley1982meaning}:

$$\text{AUROC} = \sum_{i=1}^{N} \frac{(FPR_{i} - FPR_{i-1}) \times (\text{Recall}_n - \text{Recall}_{n-1})}{2}$$

Similarly, this metric measures performance between 0 and 1, but this time with .5 indicating a perfectly random prediction and 1 a perfect classifier \citep{janssens2020reflection}. One caveat with AUROC and ROC curves in general in the presence of imbalance is that they can be. The proposed  solution to this problem is the use of PR curves and AP as an alternate (or even primary) metric -- which is the solution employed by this paper \citet{janssens2020reflection, sofaer2019area, saito2015precision}.

The advantage of these metrics is that they are not based on a cutoff - they do not depend on an arbitrary choice of probability threshold in order to determine what is a predictive positive, but instead iterate across all possible alternatives, and explore multiple tradeoffs simultaneously\citep{davis2006relationship}. 

\subsection{Annotation budgets and Convergence}

The paper assumes an average annotating speed of 50 articles/hour by an experienced professional coder\footnote{Based on timing the data collection effort for the religious coding dataset.}, leading to three investigated annotation batch sizes, of 50, 100 and 150 articles per active learning round, with the overall budget (denoted as $B$ further) being determined empirically as the point where the classifier stops improving by a substantial margin, defined as 5\% over the set of simulations.

Determining the stopping criteria, i.e. the number of active learning iterations is done by running Monte-Carlo (MC) active learning simulations, and stopping when no improvement of the AUPR score over a window of three iterations or more. The window is set such that a slight loss of performance due to an accidental improvement of an outlier is eliminated.

\section{Where to start: The Unsupervised Step}

The unsupervised stage is constructed based on the recommendations of \citet{dor2020active} for a practical active learning approach in the contexts of both extreme imbalance and text-based data. The motivation for such a first stage is straight-forward - the lack of balance in the dataset would mean that with a random sampling strategy of any kind, the human annotator would be exposed to an overwhelming set of negatives and very few, if any, positives\footnote{On average this would result to, on average, a single article concerning religious violence and two articles concerning electoral violence for a sample of 100 articles to code} - making any machine learning task impossible (as the classifier would not have information to read). Instead, the paper tries to extract a "core" dataset that attempts to preserve as much as possible of the positives while reducing the amount of negatives to a more manageable level - following \citet{dor2020active}'s "practical imbalance" suggestion.

This can be achieved in multiple ways - \citet{dor2020active} suggests a simple keyword search, using a subset of 1--3 keywords in their approach. However this would not be particularly feasible here, especially given the type of data generating process and the expected subjects that we may want to mine for in text-based event datasets. This is for three reasons: 
\begin{itemize}
\item Firstly, is quite common that news articles will contain either backgrounders or contextual information as decorators, especially when referring to more remote locales - eg., articles written around election periods in low-media attention countries will contain a boilerplate sentence indicating elections are forthcoming (or recently occurred)\footnote{See appendix for examples.}. 
\item Secondly, many of the subjects that one would be interested in mining for are topics of high social importance (such as religion and electoral processes), which frequently come up in news articles even when they are only indirectly connected with the event, e.g. a priest or community leader may be the spokesperson of a group or the trusted eyewitness of an event, even when that event was not in the class of events of interest for us\footnote{See appendix for example}.
\item Thirdly, most of the base datasets that can be used, such as UCDP GED or ACLED are already created based on a simple keyword search \citep{croicu2015ucdp, raleigh2010introducing}, increasing the risk of bias through double filtering.
\end{itemize}

These three issues, uncorrected, would lead to extraction of substantially more false positives (and thus fewer true positives for the human to annotate in the first round), which in turn would lead to biases, especially in those contexts where the three patterns described above are more common. For these reasons, a slightly more complex yet still computationally simple approach was employed, using parts-of-speech (POS) tagging techniques to attempt to link the subjects and concepts of interest (religion/elections) to the actions of interest (violence).

To accomplish this, two dictionaries have been created - one of verbs and one of nouns of interest, starting from a few key concepts and then mining the Princeton WordNet \citep{fellbaum2010wordnet}, a large dictionary relating words based on their hierarchical semantic space, for all related terms (hypernyms, hyponyms and synonyms). The unlabelled text in each article as well as the dictionaries were then stemmed (i.e. reduced to the its root form). Then, for each article, all identified noun-verb pairs, both within-sentence and across-sentence were identified. Then, a score was computed using the following formula:

$$ S_{art} = \sum_{k=1}^{DN} \sigma(N_k)e^{\lambda \min((pos(N_k)-pos(V_{DV}))^2)} $$

where $\min((pos(N_k)-pos(V_{DV}))^2)$ is the squared minimum distance (in number of sentences) between a verb and a noun, $sigma$ is a score assigned (by the researcher) to each of the nouns in the dictionary, and $e^\lambda$ is an exponential decay function.

The reasoning behind using this scoring function is rather simple - it is more likely that a text contains a true positive if those nouns describing the topic of interest are located close to verbs describing the action of interest, i.e. in the same sentence or in a nearby sentence. The decay function attempts to coarsely but plausibly model this likelihood using a simple functional form\footnote{Using such functions have been shown to be effective by e.g. \citet{lodhi2002text}}.

\begin{figure}[tb]
\includegraphics[width=9.9cm]{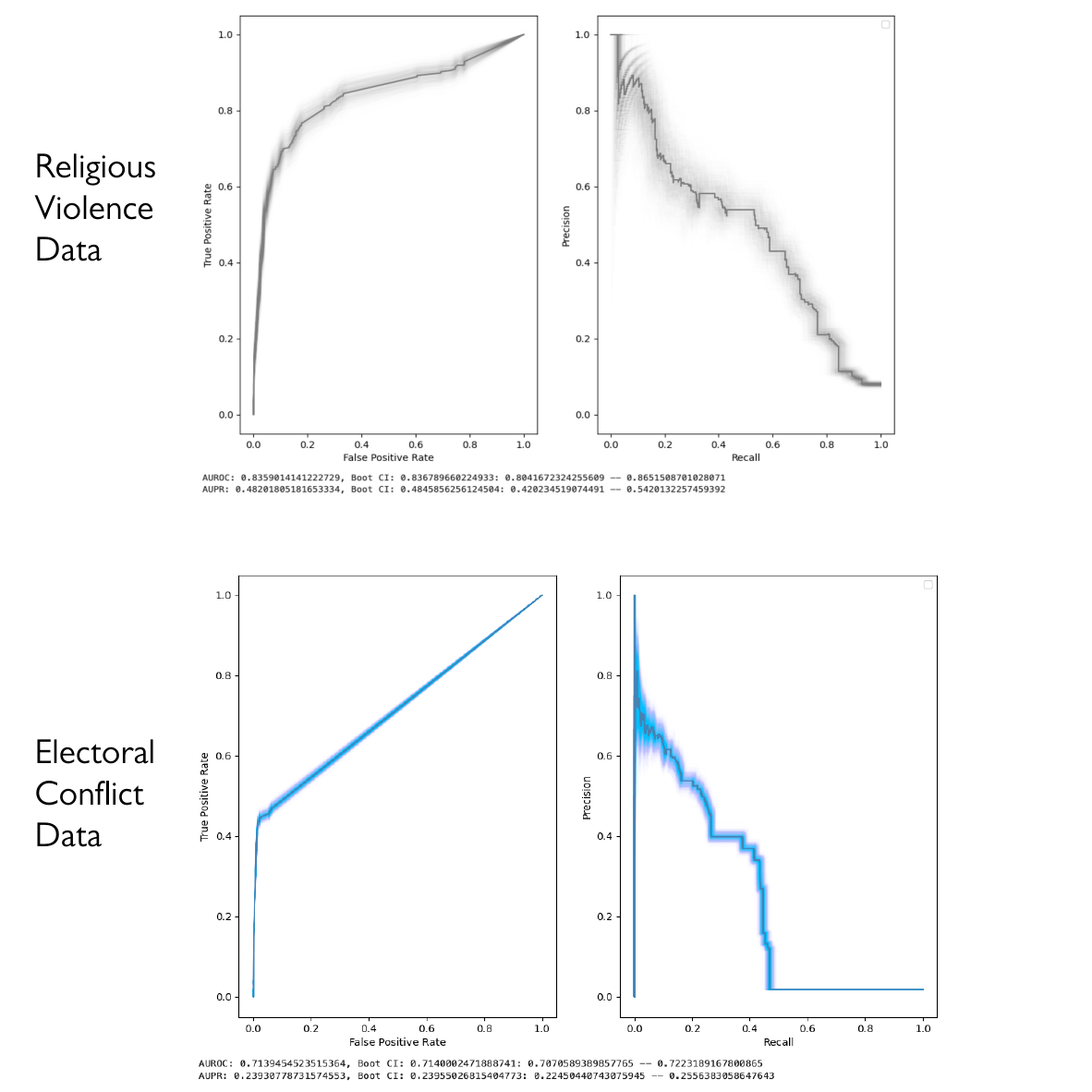}	
\caption{Receiver-Operator Curve (ROC) and Area Under the Precision-Recall Curve (AUPR), including bootstrapped confidence intervals for the purely unsupervised approach for the two datasets - top (blue) for religious violence, bottom (gray) for political (electoral) violence.} \label{fig:unsuper}
\end{figure}

Figure \ref{fig:unsuper} shows the performance of this purely unsupervised step in terms of two standard composite performance metrics for classification tasks - the receiver operating characteristic (ROC) and the precision-recall curve (PR)\footnote{These were computed against the raw $S_{art}$ scores.} \citep{tibshirani}. For both datasets, performance of this unsupervised step is relatively modest, with an average precision (area under the PR curve) of .48 for religious violence and .24 for electoral violence and an area under the ROC curve (AUROC) of .835 for religious violence and .714 for electoral violence. As such, in order to be able to extract 50\% of the true positives of religious violence events (50\% recall), over half of the resulting labeled positives would be false positives; conversely, for 90\% recall, over 90\% of the events labeled positive would be false positives. Performance is worse for the electoral violence dataset - the unsupervised classifier is unable to extract more than approximately 48\% of the true positives without labeling the entire dataset as positive (100\% false positive rate).


The coarseness of this approach is irrelevant for the task - this is simply a pre-filter, a seed for the active learning classification model, aiming to improve balance - i.e. keep as many of the true positives as possible, filtering out the (most) clear negatives from the corpus and giving some sense of the probability (to be used as a prior) of a piece of text containing a positive observation. 

Further, as this is a preliminary step, it has been designed with the express purpose of maximising recall (i.e. the ability to extract as many true positives as possible from the corpus, and lose as few of them as possible), even at the cost of precision (i.e. the inclusion of false positives in this stage is acceptable). Thus, the method is clearly sufficient for this "coarse" first sifting stage.

\section{How to learn : Picking a model}

Two types of machine learning models suitable for text classification were considered to be employed in conjunction with the active learning approach:
\begin{enumerate}
    \item a now-classical approach, based on word- and document- vector in a classical regression setting. In this approach, text (at the level of individual words or whole documents) is represented as numeric vectors, in a large multi-dimensional space (of usually 100-1000 dimensions) referred to as the embedding space \citep{church2017word2vec, bojanowski2017enriching}. The embedding space is trained using relatively simple methods, primarily simple feed-forward neural networks, designed in such a way as to be encoding the relationships between pairs of words located in the near vicinity to one-another (skip-grams) \citep{church2017word2vec, bojanowski2017enriching}. These numerical vectors are then aggregated for each article (event) and treated as features (predictors) for a classical (shallow) but state of the art machine learning model, a gradient boosting tree ensemble implemented using LightGBM \citep{ke2017lightgbm}. Such approaches and models have had significant success in domains related to political science, including conflict and contention topics, but are now considered for the most part obsolete in machine learning research and NLP research\citep{rodriguez2022word, rodman2020timely}.
    \item A contemporary approach, based on exploiting the capabilities of pre-trained, large language model (LLM) such as BERT \citep{devlin2018bert} or GPT \citep{radford2018improving}. These are deep (multi-layered) neural networks, pre-trained on a generalized language task\footnote{such as predicting the next word (in GPT) or a missing word (in BERT) for a given sequence.}, usually on extremely large corpora (of 1+ billion words). These models usually work by building a contextual representations of text at word level -- they do this by employing a series of multi-headed attention blocks -- small networks designed to relate each word with (partial aspects) of their semantic contexts across the corpus\citep{vaswani2017attention}. These general purpose models are extremely efficient at natural language learning and can then be repurposed (through a transfer learning approach referred to as fine tuning, where part of the architecture of the model is altered, and then the weights of the models are further trained) for other tasks where natural language is employed with relatively little effort and little additional training material being required \citep{brown2020language}. These models have recently been shown to perform well in contexts of interest, e.g. with general political science tasks such as classification of political manifestos and speeches \citet{laurer2022less} as well as with conflict related domains such as improving on conflict event extraction and terrorist event classification \citep{hu2022conflibert}. 

    \citet{laurer2022less} show these types of models, even when trained on general text, can be used to forecast with relatively little fine-tuning a transfer-learning approach.\footnote{The Natural Language Inference (NLI) approach in \citet{laurer2022less}, where a context (in the form of a hypothesis) is provided together with the text at training time, was considered but not used -- since the corpus employed is already passed through one classification round, the extraction of violent events, an NLI approach trained on a general corpus of hypotheses and prompts would only overfit on the general conflict pattern already extracting, and coding specialized hypotheses would be equivalent in time with manually coding the dataset.}. Furthermore, there now exist classes of models specifically pre-trained on the types of texts and contexts related to conflict information extraction and forecasting such as ConfliBERT \citet{hu2022conflibert} which also outperform the state of the art.

    Further, two sub-types of these models can be distinguished:
    \begin{enumerate}
        \item \textit{encoder-based} networks such as BERT, where the corpus is distilled into a smaller-dimensional semantic space explicitly targeted at capturing word-context relationships (through an architecture referred to as an encoder).
        \item \textit{decoder-based} approaches, such as GPT, where context representation is not done explicitly but implicitly (there is no encoding in a smaller-dimensional semantic space), and semantic relationships are only captured against elements coming earlier in the text (the model is prevented from reading text ahead of the current word through a process called masking).\citep{radford2018improving} 
    \end{enumerate}
    
    While theoretically, encoder-based models should perform better for classification tasks, as they explicitly distill contexts into smaller dimensional semantic spaces which should aid classification that is essentially a reduction of that space to the dimension of the number of classes - the larger architectures of decoder-based approaches and much higher numbers of parameters seem to overcome that advantage, and thus both of these are considered below \citep{naveed2023comprehensive}.
\end{enumerate}

\begin{figure}[tb]
\includegraphics[width=9.9cm]{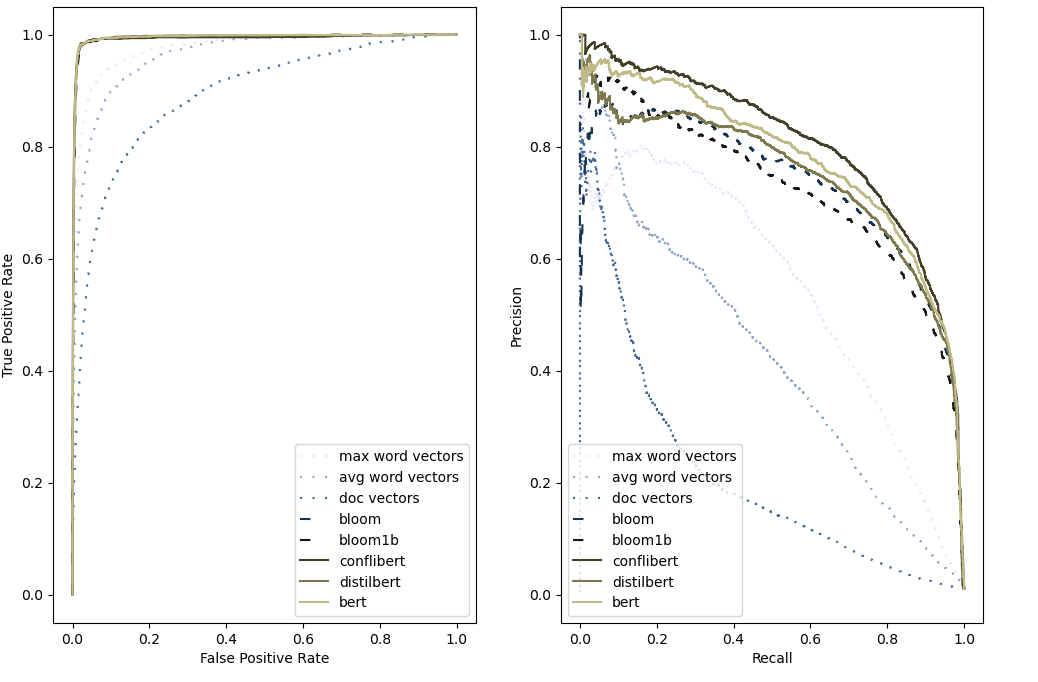}	
\caption{Cross-validation experiment (with complete data) between three classes of natural-language classification models (8 models) for extracting religious violence events. Left pane - ROC curves (note high imbalance), Right pane - Precision Recall Curve. In both cases higher curves indicate better performance. Blue models are classical (shallow learning) approaches; dashed brown lines are decoder-only LLMs; solid brown lines are encoder-only LLMs.} \label{fig:cv}
\end{figure}

Given the large choice available, to ground the choice of base language modelling approach, two simple cross-validation experiments have been run. These exploit the fact that both datasets employed are gold standard and contain completely labeled information.

This are run as following -- both the religious violence data and the electoral violence data are split into 5 folds. Four folds are used for training (fine-tuning LLMs for the deep learning approach and training word vector embeddings and corresponding regressions for the shallow learning approach), and the classes of the fifth partition is predicted. This is done in a round-robin approach.

Models are selected from each class of models described above -- Word2Vec word vectors  and Doc2Vec document-level vectors coupled with a state-of-the-art LightGBM boosted tree classification model for the shallow learning approach, two fine-tuned general-purpose BERT models (the smaller Distilbert with 66 million parameters, and base BERT with 110 million parameters, trained on approximately 100 GB of text) and the conflict-specific ConfliBERT (uncased, fully pre-trained, 110) for encoder-based LLMs, and two open-source decoder-only LLMs (a smaller one, Bloom-560m with 560 million parameters and a larger one, Bloom-1b with 1 billion parameters, both trained on approximately 1.5TB of text) for decoder-based LLMs\footnote{Bloom is a state of the art open-source decoder-based LLM, reflecting current advances in architecture on par with closed-source approaches such as GPT-3.5.}.

The same cross-validation partitioning is used across all models under investigation (thus all models are exposed to the same data). All LLMs are fine-tuned for binary classification tasks -- their last layer is replaced with a fully connected layer with sigmoid activation. Training using the cross-validation train partition (80\% of the data) is then performed on top of the existing (pre-trained) weights and biases, and out-of-sample predictions are done for the remaining 20\%. Five models are then trained, one for each test partition, and results aggregated.

\begin{table}[h]
    \centering
    \begin{tabular}{|l|l|l|l|}
    \hline
    Method Class & Method & Average precision & Area Under ROC-curve \\
    \hline
    Encoder LLM & ConfliBERT-scr-uncased & \textbf{0.8051} & 0.9938 \\
    Encoder LLM & Base BERT-base-uncased & 0.7789 & \textbf{0.9955} \\
    Encoder LLM & Distilbert-uncased & 0.7466 & 0.9950 \\
    Decoder LLM & Bloom-750M & 0.7318 & 0.9949 \\
    Decoder LLM & Bloom-1B & 0.7229 & 0.9941 \\
    Shallow & LightGBM, max word-vectors & 0.5450 & 0.9765 \\
    Shallow & LightGBM mean word-vectors & 0.4250 & 0.9622 \\
    Shallow & LightGBM document vectors & \textit{0.2123} & \textit{0.8922} \\
    \hline
    \end{tabular}
    \caption{Performance metrics for the evaluated models in a cross-validation (gold standard data) scenario, without active learning, corresponding to expected best performance. Note that AUROC is artificially high for all models due to very high class imbalance (99\%).}
    \label{tbl:cv}
\end{table}

Results for the the performance of classifiers of various type discussed above in extracting religious violence events are presented in figure \ref{fig:cv} and \ref{tbl:cv}\footnote{Results for electoral violence follow the same pattern and are presented in the web appendix.}. These can be interpreted as the "best-case performance" for those classifiers for the task in the presence of gold-standard complete date, i.e. this is the maximum benchmark that the active learning approach can theoretically reach.

All encoder-only LLMs (slightly) outperform decoder-only LLMs, even where the encoder-only network is over 10 times less complex in both parameter space, amounts of pre-training data and computational complexity at both train and inference time. 

In turn, all neural network-based LLMs outperform classical approaches significantly, at high recall levels LLMs predicting over an order of magnitude fewer false positives.

Between encoder-only LLMs, ConfliBERT, pre-trained specifically on the type of text most commonly used in conflict research, slightly outperforms the more general models, but all three models examined performed excellently, including the much smaller and computationally inexpensive DistilBERT.

Given the results, the active learning effort will make use of encoder-based LLMs - further experimenting with ConfliBERT and DistilBERT. Carrying on experiments with both models is advantageous for two reasons -- on one hand, avoids any  pitfalls from the potential overfitting on conflict contexts originating from the pre-training of a conflict-specific LLM outside the control of this paper; on the other, allows for experimenting with a computationally simpler and slimmer model that is more tractable for practical use with the resources typically available to researchers in the field.

\section{How to employ the human : The Active Learning Rounds}

Setting up an active learning approach needs to take into account the standard practices for human annotation that are prevalent in the field, so as to guarantee that human resources are available for practical use of the method beyond simple experimentation. This is even more important as most data collection efforts such as UCDP or ACLED have long established practices when it comes to data collection and annotation routines \citep{sundberg2013introducing, raleigh2010introducing}. Thus, the design of the active learning method must be centered around these paradigms. The most important of these, and the most consequential for the design of the approach is batching -- human annotators are not available constantly, and it is much easier to present a (larger) batch of data to an annotator than to ask to annotate one article at a time at the rhythm of the machine learning algorithm\footnote{This is even more important as inference in transformer-based LLMs is relatively slow, and given the large data unbalance, even a well trained classifier will require quite a lot of time to select a suitable new candidate for classification.}.

As such, batch annotation will be used for each round of annotation, with the overall annotation budget for human coding thus being the (annotation) batch size $b$ multiplied with the number of active learning rounds $r$. 

The active learning process starts with the human annotator\footnote{oracle in active learning terminology} annotating 200 articles selected from the seed data (the corpus as pre-processed in the unsupervised step). The 200 articles are explicitly selected in such a way as to maximize potential sample balance (i.e. to give the machine learning model a reasonable sample of both positives and negatives to learn from). Since the overall balance of the data is not known a priori, and imbalance is substantial (at 1--2\% positives), canonical approaches do not work. 

Instead, an experimentally determined polynomial sample weighing function $W_{art}=s_{art}^4$  was determined through Monte Carlo trials on the two complete datasets, resulting in a function that would result in a starting seed as close to balance as possible while being as stable as possible and generalize across the two domains for which data was available (electoral violence and religious violence). The scoring function is then used as the weighing function to perform Monte Carlo weighted sampling, with the sampling function being:

$$Pr(annotation|{art}) = \frac{s_{art}^4}{\sum{s_{art}^4}}$$.

These are then manually classified by a human annotator, and these become the first batch of training data for the first round $r=0$ of fine-tuning of the active learning classifier. These are then used to fine-tune the selected deep learning model to create a classifier trained at round 0 of active learning ($C_0$).

After this first round, for each round $r>0$:
\begin{enumerate}
    \item The trained classifier $C$ from round $r-1$ predicts the probability of the article being a positive observation (e.g., for our test datasets, religious violence or electoral violence). For computational complexity, given the extreme nature of unbalance, this is only done for those articles in the dataset containing a score $s_{art}>0$.
    
    \item From this dataset of articles, a new sample $D_{b,r}$ containing $b$ new unlabeled articles for annotation is taken through a weighted sampler with weights based on two-class (binary) entropy ($W_h$) \citep{dagan1995committee, tibshirani} given the probabilities predicted by the classifier:

    $$W_{art} = - [\hat{p}_{art|C_{r-1}} \log_2(\hat{p}_{art|C_{r-1}}) + (1 - \hat{p}_{art|C_{r-1}}) \log_2(1 - \hat{p}_{art|C_{r-1}})].$$

    This is referred to in the active learning literature as the querying strategy \citep{dor2020active, burr2009, lin2018} and is the selection mechanism employed to identify appropriate unlabeled instances for human annotation \citep{burr2009}. Intuitively, the objective of $W_h$ is straightforward: given that there are constrained resources available for manual annotation (this being the primary rationale for employing active learning), the primary concern is optimizing the use of these resources to most significantly enhance the classifier's performance. In essence, this equates to identifying the samples nearest the decision boundary of the classifier, i.e., those about which the classifier is most uncertain \citep{burr2009}. For a balanced dataset, the criterion often simplifies to selecting instances where the classifier outputs a predicted probability nearest to .5 \citep{burr2009}.
    
    However, when faced with class imbalance -- where the number of labels for one class (e.g. battle events concerning religious conflict) are greatly outnumbered by the number of events of the other class (e.g. all other kinds of conflict) -- selecting an appropriate querying strategy is much more challenging. The risk in such cases is that the human annotator is presented with only the (easy to predict) majority class observations, leading to no improvement made through human labelling (as predicting the majority class is trivial). 
    
    Indeed, finding an appropriate querying strategy is an open research question in computer science literature \citep{lin2018, ertekin2007learning}. Specifically, the challenge lies in the heightened difficulty of isolating unlabeled instances of the minority class relative to those drawn from the majority class \citep{lin2018, ertekin2007learning}. This is further complicated in such cases where the distribution of the classes cannot be determined in advance, as that implies that there is no way to exploit such distributional assumptions \citep{ertekin2007learning}. 

    The choice of binary cross-entropy for this task was made following \citet{dor2020active}'s suggestions with the caveat that most of the more complex methods proposed under-perform in the context of the extreme data unbalance present, and collapse, presenting only the 0-class to the coder.

    \item This newly selected batch ($D_{b,r}$) of $b$ articles are then coded by the human annotator, and removed from the dataset of unlabeled articles. As the datasets used for the experiments presented in the paper are completely coded, this process is simulated in the experiments. However, a practical approach was developed using UniversalDataTool for practical use -- the classifier is stopped and its weights and biases saved, the selected articles outputted and stored on a centralized server and sent to the coder that can then load them in a simple coding interface for annotation.

    \item A new classifier $C_r$ round is trained by taking the weights and biases of the previous round classifier $C_r$ and further fine-tuning it with the articles in the current batch $D_{b,r}$.
    
    The round is treated as a (single) new training epoch for the classification model, i.e. it is done as a single training pass over the entirety of the newly annotated data, starting from the weights generated in the previous training round. Per the standard approach in BERT models (and deep learning models in general), articles are batched, with a back-propagation run for each run\footnote{This (technical) deep-learning article batching is separate from the annotation batch - i.e. the coding batch of 50--150 articles are split into 7--19 sub-batches of 8 articles. This size is used for computational (memory) considerations, especially at inference time -- article lengths are quite long, leading to quite large representations in memory (each article is represented as a sequence of 500 tokens).}.

    \item The process is stopped when either improvements in terms of average precision (AP/AUPR) cease -- defined as mean improvements in AUPR being, for each round, lower than a conventional threshold (set here at 1\%) \citep{tibshirani} -- or when a maximum number of iterations have been exceeded. Evaluation is done fully out-of-sample -- and for this purpose, since the datasets under evaluation are complete (gold standard) -- 20\% of the data has been reserved specifically for evaluation and excluded from the active learning loop altogether. When applying this method in practice -- where such gold standard data is not available, in order to insure face validity of the method, as well as to avoid any potential overfitting concerns by over-training the classifier, this could be replaced with a small amount of manually collected positives and negative cases, manually reserved from the dataset, in order to evaluate the performance of each round $r$ against\footnote{To maintain decent balance, using the same $s_{art}^4$ sampling weights to select a number of 50-100 cases manually for validation has proven to be effective empirically when transferring this method to other domains such as attacks on transportation infrastructure or attacks on health services.}.
\end{enumerate}

\section{Results}

\begin{figure}[ht!]
    \centering
    \subfloat[\centering Religious violence (AP above, AUROC below)]{{\includegraphics[width=7.5cm, height=10cm]{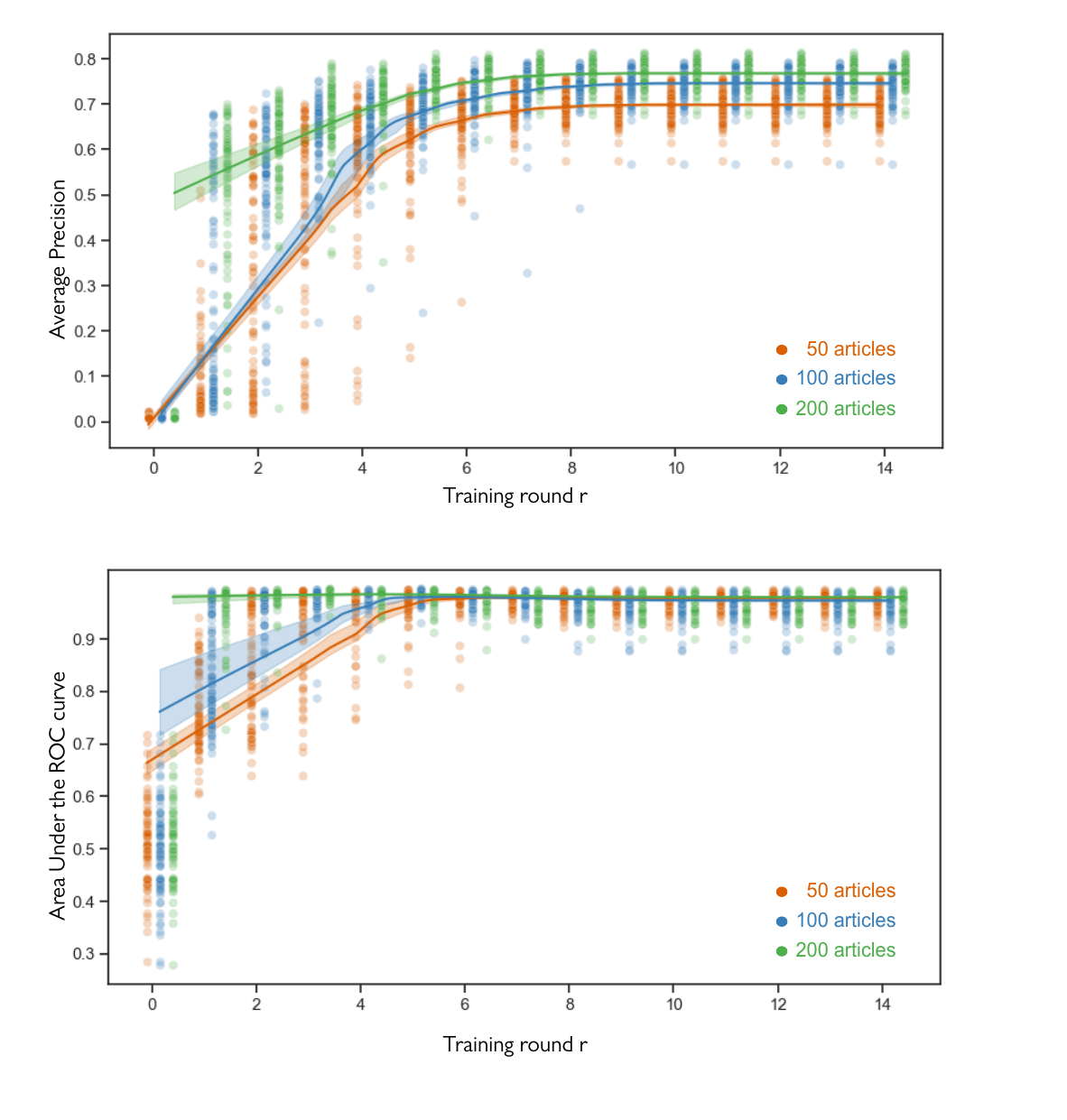} }}%
    \qquad
    \subfloat[\centering Electoral violence (AP above, AUROC below)]{{\includegraphics[width=7.5cm, height=10cm]{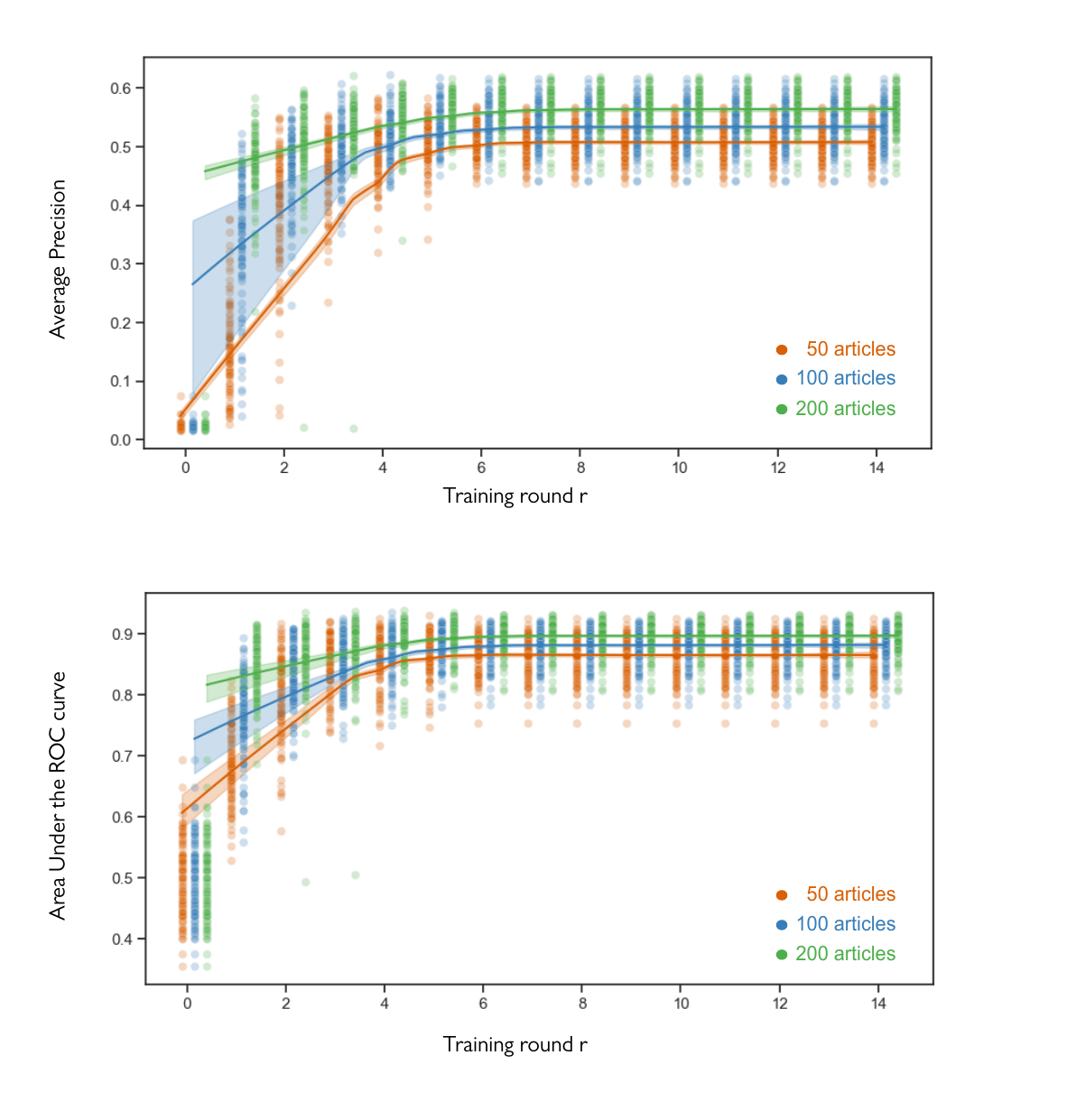} }}%
    \caption{Average Precision (AP) and Area Under the Receiver Operator Curves (AUROC) across active learning rounds for three batch sizes (50, 100 and 200 articles) on the two complete datasets: religious and electoral violence. Learning rate for all experiments : $5 \times 10^{-5}$}%
    \label{fig:resultsm1}%
\end{figure}

The active learning experiment described in the section above was simulated with three specifications for the three annotation budgets described above -- with batch sizes $b$ of 50, 100 and 200 articles being simulated as sampled from the previous iteration of the classifier, sent to the human annotator for annotation, and returned for continued fine-tuning of ConfliBERT. 

In order to get an understanding of model uncertainty, and in order to avoid drawing conclusions that are potentially based on overfitting on outliers present in a single dataset, 100 Monte Carlo simulations were performed for each specification and for each dataset (resulting in 300 models, run for 15 active training rounds each)\footnote{The size of each experiment was driven by computational considerations, since each run takes approximately 4 GPU-hours to run.}. In each such experiment, a different test partition of 20\% of all articles is sampled and reserved for out-of-sample evaluation (maintaining the traditional proportions from classical cross-validation) and never shown to the classifier at any round except final prediction. 

The remainder 80\% serves as the (semi-supervised) training pool from which the training data is drawn through the active learning approach. From this pool, batches are drawn for (simulated) human annotation. This  Monte Carlo based approach, inspired by \citet{carsey2013mc}, was chosen instead of classical cross-validation as there are two sources of potential overfitting -- the data points that are present in the train/test split as well as the data points that are extracted for each active learning run. Moreover, since only a very small subset of data is used to train the classifier at round $r$ (50--200 articles), the risk of potential overfitting is higher. With this setup, where both the test and train data is repeatedly resampled, this risk is mitigated further. The same procedure was repeated twice -- once for the electoral violence and once for the religious violence dataset.

\subsection{Convergence and active learning rounds}

Figure \ref{fig:resultsm1} show the results of these 600 active learning experiments, 300 for each of the two datasets -- i.e. the religious and electoral violence datasets. Each of these experiments were run across 15 active learning rounds ($r$), with the 100 each for each of the three batch sizes\footnote{Full results and further specifications are presented in the web appendix.}. 

Across all specifications of religious violence, convergence occurs around round 6 where the mean improvement in AP across all simulations between rounds 5--6 in the AP domain is 3.01\%. 64\% of all classifiers exhibit a performance increase in the AP domain of over 1\% and 19.5\% exhibit a performance increase of over 5\%. This drops significantly at round 7, with mean performance increase of 0.44\% (dropping further to a mean performance decrease of -0.0001\% in round 8), with only less than 12\% of all classifiers exhibiting an increase in performance of over 5\% (tapering off to 4\% of all classifiers in round 8).

For electoral violence, results are nearly identical, convergence occurring at approximately the same point, with a mean improvement of 0.77\% in the AP domain between rounds 5--6, becoming slightly negative (-0.3\%) at the next round. 9.6\% of the classifiers improve more than 5\% and 41\% of more than 1\% between rounds 5 and 6 in the AP domain. Similarly, this drops for the next round to 3\% and respectively 0.6\% for rounds 6--7. 

Increases beyond these rounds are marginal, indicating the classifiers (in both domains) no longer learn any improvement after about round 7.

Overall performance is, moreover, excellent. For the religious violence dataset, median AP is 0.7645 (95\% quantile CI : 0.6915 -- 0.8121), and median AUROC is 0.9812 (95\% quantile CI : 0.9533 -- 0.9916). For political violence, across all experiments, median AP is 0.5486 (95\% quantile CI : 0.4716 -- 0.6118) median AUROC is 0.8908 (95\% quantile CI : 0.8211 -- 0.9249). These results, where the ConfliBERT classifier is trained on 700 -- 1,400 articles are extremely close to the best-case, gold standard cross-validation baseline, where training was done on 193,114 articles and where the values for the entire dataset is known a priori.

The performance of the resulting active learning classifier is practically indistinguishable from the gold standard cross-validation approach : the observed increase in median error between the active learning approach against the fully-annotated, in AP domain is 0.04, and median error increased in AUROC domain by 0.012. 

In practice this extrapolates to fewer than ten supplementary false positives over the whole dataset, when using the active learning approach, versus when having the whole hand-coded dataset of 193,114 articles\footnote{Note that performance in AP domain is even higher than the cross-validation score in 4.8\% of the Monte Carlo experiments -- this is due to the randomness of the test dataset, resulting from its repeated resampling in each MC experiment.}.

This level of performance, despite low amounts of training data  confirms results from other domains -- they perform well in few-shots contexts -- and increasing the amount of data in training beyond a relatively low threshold does not equate in significant further improvement, making them particularly suitable for active learning approaches \citep{ahmed2022few, song2023llm}.

However, one caveat, for the first few rounds, the classifier performance increases extremely fast -- at least in the current setup, these are not zero-shot approaches, and should be trained at least with a few rounds for decent performance\footnote{Although a prompting-based strategy may improve zero-shot performance, developing such an approach that would generalize well was not straightforward in the limited experimentation made; a more traditional fine-tuning approach immediately yielded results.}

But what does this level of performance mean for practical use? In practical terms, these levels are at or exceed the state of the art for related problems in state-of-the art broader political science (e.g. results are generally on par with those presented \citet{laurer2022less}, despite the much higher imbalance in the two datasets explored in this paper, and the much smaller human annotation budgets available).

As such, to correctly label 75\% of the 1982 articles of religious violence a median of 3203 total articles would have to be classified as positive (95\% CI: 2984 ---4381) and 190,134 articles would be eliminated, equivalent to a recall of the false category of 99.21\%. A higher desired recall rate does not affect the performance of the classifier much, to extract 90\% of the same number of articles would require 8273 articles to be classified (95\% CI: 4371--20697), equivalent to a recall of the false category of 97\% (95\% CI: 90.95\% -- 96.70\%). This is a substantial improvement to what was possible even a few years ago, showing significant improvement to the 50--70\% recall false target in \citet{croicu2015improving}, and performance above that of \cite{laurer2022less}, having less unbalanced targets\footnote{However, such comparisons are only for illustration and very coarse, since the targets of those classifiers were not the same as this one}.

For the political violence dataset, results are somewhat weaker. To correctly label 75\% of the 4016 articles of political violence a median of 39067 total articles would have to be classified as positive (95\% CI: 19479 - 71194) with a mean of 134,085 articles would be eliminated, equivalent to a median recall of the false category of 84.05\% (i.e. the filtering out, on median of 84\% of true negatives). This lower performance, while still appropriate for many tasks (e.g. forecasting or filtering before a final manual pass), can be explained by a number of issues. Based on an investigation of a fixed 20\% sample of the data -- classification error comes from four sources:
\begin{enumerate}
    \item Human annotation is conducted solely or mainly based on context and area expertise external to the text in the corpus, e.g. based on the knowledge that two ethnic groups are in conflict with one-another during an electoral campaign, or that it is an electoral period. This is much more common in the electoral violence dataset, explaining the substantially lower performance of the electoral violence classifier. Mitigating this could be done by including (known) structural data (such as e.g. whether a campaign is occurring, or whether an ethnic group is tied to campaigning). This could be done with ease in classical data mining approaches, where that data is known, or, for predictive analytics, could be gleamed from multiple passes over the data to identify related actors, with a first pass done with e.g. the approach in \citet{parolin2021come}.
    \item Multiple events being amalgamated and presented in the same news-text. While this most frequently works well, with very brief mentions of events scattered around articles, the process fails. A segmentation approach such drawn from the image segmentation literature such as in \citep{minaee2021image} may work well to mitigate this issue and further improve the effort. 
    \item Very rare events, e.g. attacks against very small or rarely attacked denominations (e.g. Buddhists) tend to be harder to classify by the machine, especially when no example was encountered in the training data. Running multiple chains of active learning and ensembling the results through a simple arithmetic mean of the resulting probabilities would mitigate this error.
    \item Further, a number of human coding errors were captured, e.g. when a planned attack against a church was foiled and the attackers-to-be killed by police, or when electoral violence was coded nearly the halfway point between two elections. These were not revised in the original gold-standard datasets, meaning that performance above 91\% recall (the equivalent Cohen coder inter-reliability score obtained) is not even theoretically possible, as humans could not agree on the code above that level.
\end{enumerate}

Resolving these issues can be done in multiple ways -- a final manual pass on the classified data, which becomes much more tractable with elimination of 80--95\% of the true negatives, ensembling of multiple chains of active learners in a similar way as proposed by \citet{dor2020active} or \citet{croicu2015improving} etc.. 

Indeed, ensembling does significantly improve performance for both the religious and political violence classification : from an AP of 0.7208 (95\% CI: 0.6551 -- 0.7517) for a normal single active learning run over 7 iterations to 0.7799 (95\% CI: 0.7606 -- 0.7991) with an ensemble 5 active learning chains and 0.7909 (95\% CI: 0.7766 -- 0.8031) when 10 active learning chains are ensembled\footnote{All results obtained from 1000 simulations from 50 active learning runs trained for 7 rounds. See the web appendix for detail and a further discussion}. For electoral violence a similar improvement occurs, from 0.5266 (95\% CI : 0.4594 -- 0.5615) for a single active learning run to 0.5946 (95\% CI: 0.5642 --  0.6216) for 5 runs ensembled to 0.6100 (95\% CI : 0.5909 -- 0.6289) for 10 runs ensembled. Note, however, that ensembling multiple active learning rounds will require substantially more human annotation effort.

\subsection{Choice of batch sizes}

\begin{table}[]
    \centering

\begin{tabular}{llrrrrrrrr}
\toprule
 &  & \multicolumn{2}{r}{AP Religion} & \multicolumn{2}{r}{AUROC Religion} & \multicolumn{2}{r}{AP Electoral} & \multicolumn{2}{r}{AUROC Electoral} \\
 &  & mean & std & mean & std & mean & std & mean & std \\
Round & Batch Size &  &  &  &  &  &  &  &  \\
\midrule
\multirow[c]{3}{*}{3} & 50 & \textit{0.5515} & \textit{0.1511} & \textit{0.9530} & \textit{0.0505} & \textit{0.4656} & 0.0586 & \textit{0.8523} & 0.0444 \\
 & 100 & 0.6529 & 0.0599 & 0.9775 & 0.0157 & 0.4990 & 0.0469 & 0.8576 & 0.0419 \\
 & 200 & 0.7056 & 0.0492 & 0.9797 & 0.0177 & 0.5251 & \textit{0.0634} & 0.8758 & \textit{0.0540} \\
\multirow[c]{3}{*}{5} & 50 & 0.6577 & 0.0705 & 0.9738 & 0.0284 & 0.5042 & 0.0391 & 0.8662 & 0.0348 \\
 & 100 & 0.7222 & 0.0385 & 0.9797 & 0.0125 & 0.5337 & 0.0338 & 0.8816 & 0.0275 \\
 & 200 & 0.7621 & 0.0340 & \textbf{0.9819} & 0.0148 & 0.5589 & 0.0293 & \textbf{0.8954} & 0.0258 \\
\multirow[c]{3}{*}{7} & 50 & 0.6980 & 0.0381 & 0.9750 & 0.0126 & 0.5074 & 0.0284 & 0.8640 & 0.0317 \\
 & 100 & 0.7497 & 0.0300 & 0.9786 & 0.0098 & 0.5331 & 0.0359 & 0.8790 & 0.0271 \\
 & 200 & \textbf{0.7725} & 0.0304 & 0.9793 & 0.0092 & \textbf{0.5614} & 0.0330 & 0.8942 & \textbf{0.0251} \\
\multirow[c]{3}{*}{9} & 50 & 0.6989 & 0.0360 & 0.9729 & 0.0129 & 0.5073 & \textbf{0.0284} & 0.8639 & 0.0316 \\
 & 100 & 0.7488 & 0.0301 & 0.9784 & 0.0097 & 0.5331 & 0.0359 & 0.8790 & 0.0271 \\
 & 200 & 0.7712 & 0.0308 & 0.9790 & 0.0095 & \textbf{0.5614} & 0.0330 & 0.8942 & 0.0251 \\
\multirow[c]{3}{*}{11} & 50 & 0.7000 & 0.0360 & 0.9734 & 0.0130 & 0.5073 & \textit{0.0284} & 0.8639 & 0.0316 \\
 & 100 & 0.7506 & \textbf{0.0288} & 0.9789 & \textbf{0.0092} & 0.5339 & 0.0357 & 0.8798 & 0.0267 \\
 & 200 & 0.7719 & 0.0307 & 0.9791 & 0.0093 & 0.5611 & 0.0330 & 0.8943 & 0.0252 \\
\bottomrule

\end{tabular}
    \caption{Performance of various batch sizes and selected training rounds for classifying events of religious and electoral violence in the AUROC and AP domains. Mean scores and standard deviations across MC experiments.}
    \label{tab:bss}
\end{table}
As seen in figure \ref{fig:resultsm1} and table \ref{tab:bss}, larger batches $b$ do slightly improve performance of the overall classifier applied to both the religious violence and electoral violence datasets, but this variation is not particularly high, especially at annotation budgets of over 100 articles per round. Indeed, doubling the annotation budget (and thus the cost) from 100 to 200 articles per round will provide a mean improvement of 3\% in the AP domain and of under 1\% in the AUROC domain. Such an improvement is minimal in practice and does not warrant a doubling of human labor cost.

While models are working decently well even with an annotation budget of 50 articles per round, the difference here is slightly more notable : more rounds are required for performance to converge on this budget (thus reducing savings in terms of human cost) - i.e. 11 rounds are needed for an AUPR of .7 for religious violence, and a lower overall performance (of 7\%--11\%) in the AP domain is observed.

One recommendation that stems from these results is that if annotation budgets are extremely low, further work needs to be done with the LLM hyper-parameters, in order to guarantee better tuning of signal extraction -- see also the Learning Rates section below, where significant performance gains can be made with tuning the learning rate for very low annotation budgets.

\subsection{Learning rates}

\begin{figure}[htb!]
\includegraphics[width=9.9cm]{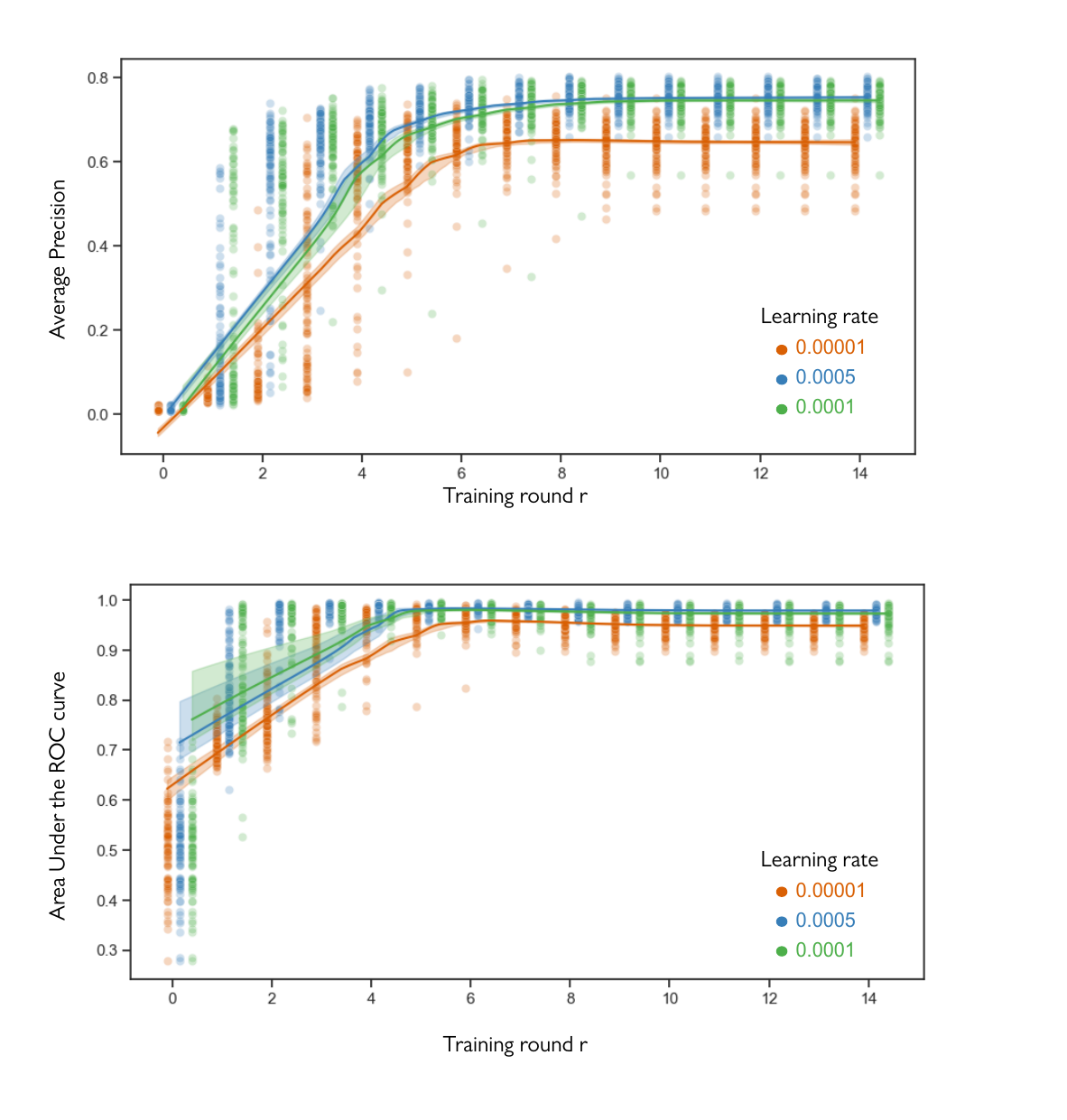}	
\caption{Impact of learning rates on the performance of the ConfliBERT-derived classifier predicting religious violence.} \label{fig:lr}
\end{figure}

Probably the most important hyper-parameter of the large language model in terms of potential influence on the performance of the active learning effort is the learning rate, i.e. the speed at which the weights of the neural network are updated with each pass over the data \citep{dor2020active, devlin2018bert, tibshirani}, i.e. how much influence each piece of data can have in one pass (one training epoch) in adjusting the model to adjust to the current classification task. 

\begin{table}[htb!]
    \centering
\begin{tabular}{llrrrr}
\toprule
 &  & \multicolumn{2}{r}{Average Precision} & \multicolumn{2}{r}{Area under ROC} \\
 &  & Mean & St.dev. & Mean & St.dev. \\
Batch size & Learning rate &  &  &  &  \\
\midrule
\multirow[t]{3}{*}{50} & 0.000010 & \textit{0.5384} & \textit{0.1641} & \textit{0.9415} & 0.0463 \\
 & 0.000050 & 0.6980 & 0.0381 & 0.9750 & 0.0126 \\
 & 0.000100 & 0.6641 & 0.1092 & 0.9575 & 0.0557 \\
\cline{1-6}
\multirow[t]{3}{*}{100} & 0.000010 & 0.6976 & 0.0300 & 0.9728 & 0.0093 \\
 & 0.000050 & 0.7497 & \textbf{0.0300} & \textbf{0.9786} & 0.0098 \\
 & 0.000100 & 0.7075 & 0.1293 & 0.9619 & 0.0567 \\
\cline{1-6}
\multirow[t]{3}{*}{200} & 0.000010 & 0.7472 & 0.0266 & 0.9747 & \textbf{0.0090} \\
 & 0.000050 & \textbf{0.7725} & 0.0304 & 0.9793 & 0.0092 \\
 & 0.000100 & 0.7436 & 0.0997 & 0.9678 & \textit{0.0337} \\
\cline{1-6}
\bottomrule
\end{tabular}
    \caption{Means and standard deviations for the two metrics, average precision and area under the receiver-operator curve for predicting religious violence using the \textbf{ConfliBERT} derived classifier across MC experiments. Active learning round 7.}
    \label{tab:my_label}
\end{table}

The design of the active learning approach, where each injection of new training data corresponds with the training epoch of the large language model, makes this even more important -- as the maximum influence that can be exerted to the model by each training round is tightly controlled by this hyperparameter.

To evaluate this influence, three different learning rates were explored across the Monte Carlo setup described above. These were determined from the last value of the BERT learning rate scheduler, and varied so that they cover the most extreme reasonable interval: \citep{devlin2018bert} : $10^{-6}, 5 \times 10^{-5}$ and $10^{-5}$.

Results are presented graphically in figure \ref{fig:lr} and in more details in the web appendix. The best performing value is the one derived from the BERT learning schedule, of $5 \times 10^{-5}$, which seems to outperform the extremes, however, the influence of the hyperparameter is modest, with improvements of approximately 7\% between the worst and best performing approaches. 

The higher learning rate performs almost as well (mean AP at round 7, for a batch size of 100 drops slightly, from 0.7497 to 0.7075, mean AUROC drops from 0.9746 to 0.9618), but with a much higher variance in the results across experiments (at round 7, standard deviation is .1293 instead of .0300 in the AP domain and .0567 instead of .0097 in the AUROC domain). This indicates that the higher learning rate is slightly too high, and the gradient descent is consistently having troubles stabilizing the weights in the area of local maxima. 

The lower learning rate performs substantially worse, with a difference in performance against the best performing learning rate being as low as 22.87\% at round 7 in the AP domain for the smallest batch size of 50. Since this delta reduces with higher batch sizes, it is indicative that at this rate, the classifier under-fits and does not have the ability to extract the full signal from each new batch of text it is presented.

\subsection{Replacing the classifier: DistilBERT}

\begin{figure}[ht!]
    \centering
    \subfloat[\centering Religious violence (AP above, AUROC below)]{{\includegraphics[width=7.5cm, height=10cm]{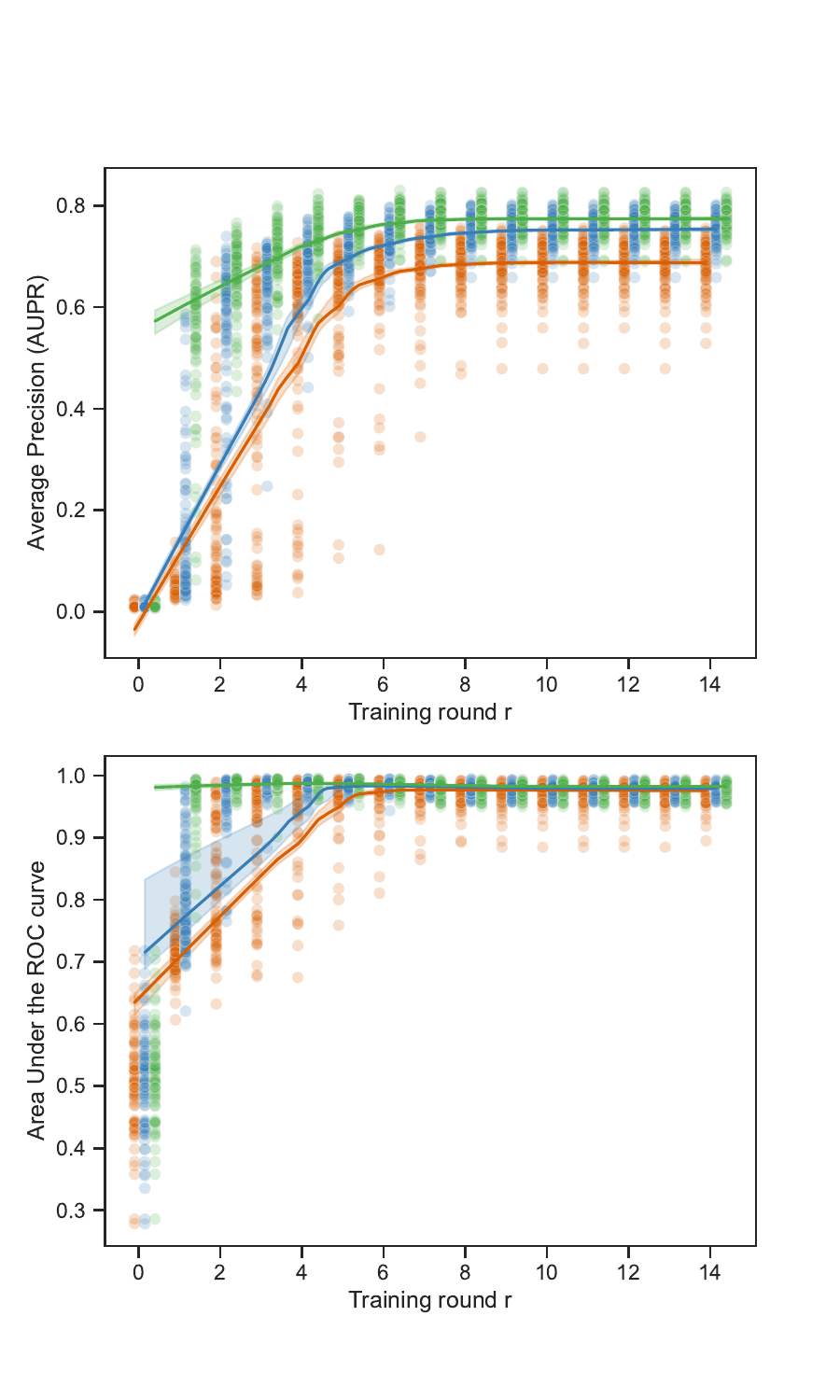} }}%
    \qquad
    \subfloat[\centering Electoral violence (AP above, AUROC below)]{{\includegraphics[width=7.5cm, height=10cm]{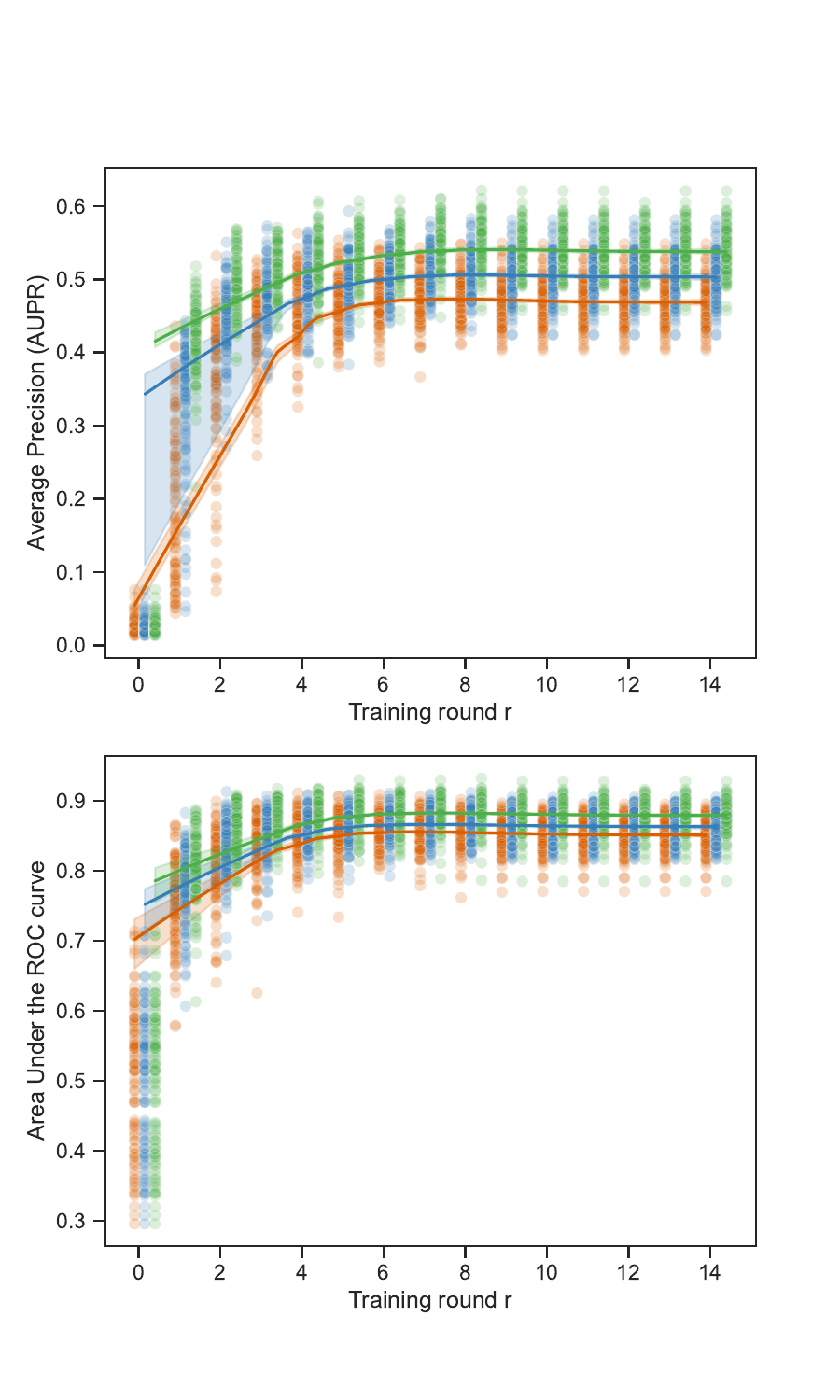} }}%
    \caption{Average Precision (AP) and Area Under the Receiver Operator Curves (AUROC) across active learning rounds for three batch sizes (50, 100 and 200 articles) on the two complete datasets: religious and electoral violence. Learning rate for all experiments : $5 \times 10^{-5}$ using DistilBERT instead of ConfliBERT.} 
    \label{fig:resultsdb}%
\end{figure}

Figure \ref{fig:resultsdb} replicates figure \ref{fig:resultsm1} with a single difference : changing the base classifier from ConfliBERT to DistilBERT, a much smaller and faster large language model that is not trained on a corpus specialized on armed conflict. Most of the results presented in the sections above do not change in any significant fashion -- findings about the number of rounds needed, the size of the batches and learning rates continue to apply with this other derivative of BERT\footnote{for detailed tables comparing the two base LLMs across training steps, learning rates etc., refer to the web appendix.}. 

Two caveat do apply -- first, with the two datasets under test, performance in both the AP and AUROC domain degrade, ceteris-paribus, by between 2\% and 10\%, depending on the set-up (batch size, number of rounds etc.), while computational complexity drops by approximately half due to the reduced number of parameters and much simpler network architecture.

Second, small-scale synthetic tests indicate ConfliBERT as a base LLM is much more overfit on the type of texts that are commonly used in the coding of violent event data -- e.g. newswire-type text and NGO/IGO reports, and shows relatively poor performance on other categories of text (e.g. narratives). Since the two datasets under consideration are based on the UCDP event corpus, a corpus very similar to the one employed by ConfliBERT, its advantage may not apply to the same extent when transfer learning to a different type of source corpora.

\section{Conclusions}

This paper introduces an active-learning method for mining existing corpora of conflict-related texts (such as the underlying corpora for UCDP GED) for new events related to micro-dynamics of conflicts, and evaluates it against two gold-standard datasets on religious and electoral violence.

Results indicate performance at or exceeding the state-of-the art in applied machine-learning techniques in the field, reducing human data collection work-loads with up-to 99\% and allowing data mining and extraction at event level even for interested individual researchers. The two datasets employed indicate good generalization potential. 

Limitations include those derived from the methodological approach coupled with the type of data under analysis -- the extreme imbalance of the type of data considered limit the application of advanced querying strategies, and the requirement for generalization preclude the inclusion of specific context-aware data together with the text in the classification effort. These remain as avenues for further research and specialization of the method.

\clearpage
\newpage

\bibliography{sample}


\end{document}